\documentclass[aps,prl,twocolumn,amsmath,amssymb,superscriptaddress,citeautoscript,longbibliography,floatfix]{revtex4-1}

\usepackage[allow-number-unit-breaks]{siunitx}
\usepackage{graphicx}
\usepackage{dcolumn}
\usepackage{bm}
\usepackage{color}
\usepackage{bbding}
\usepackage{pifont}
\newcommand{\xmark}{\ding{55}}
\usepackage{mathrsfs}
\usepackage{amsfonts}
\usepackage{multirow}
\usepackage{natbib}
\usepackage{epsf}
\usepackage{epsfig}
\usepackage{epstopdf}
\usepackage{amsmath}
\usepackage{hyperref}
\hypersetup{
    colorlinks=true,
    linkcolor=blue,
    filecolor=cyan,      
    citecolor=blue,
    urlcolor=blue,
    }

\begin{document}

\title{Contrasting Light-Induced Spin Torque in Antiferromagnetic and Altermagnetic Systems}

\author{Jian Zhou}\email{jianzhou@xjtu.edu.cn}
\affiliation{Center for Alloy Innovation and Design, State Key Laboratory for Mechanical Behavior of Materials, Xi'an Jiaotong University, Xi'an, 710049, China}
\author{Chunmei Zhang}\email{chunmeizhang@nwu.edu.cn}
\affiliation{School of Physics, Northwest University, Xi'an, 710127, China}
\affiliation{Shaanxi Key Laboratory for Theoretical Physics Frontiers, Xi'an 710127, China}

\begin{abstract}
Light-matter interaction has become one of the promising routes to manipulating various physical feature of quantum materials in an ultrafast kinetics. In this work, we focus on the nonlinear optical effects of the spintronic behavior in antiferromagnetic (AFM) and altermagnetic (AM) systems with compensated magnetic moments, which has been extensively attractive for their potential applications. With vanishing net magnetic moments, one of the main concerns is how to distinguish and disentangle AFMs and AMs in experiments, as they usually behave similarly in many susceptibility measurements. To address this challenge, we propose that linearly polarized light could trigger contrasting nonequilibrium local spin torques in these systems, unravelling hidden light-induced spintronic behaviors. In general, one could achieve light-induced spin canting in AMs, while only N\'eel vector torques in AFMs. We scrutinize and enumerate their symmetry constraints of all 122 magnetic point groups. We also adopt low energy Hamiltonian models and first-principles calculations on two representative materials to illustrate our theory. Our work provides a new perspective for the design and optimization of spintronic devices.
\end{abstract}

\maketitle
\textit{Introduction.}$-$ Spintronics serves as one of the most promising and significant fields for modern technology. While ferromagnetic materials with finite magnetic moments have been successfully used for various aspects such as magnetic storage, magnetic memory, and magnetic sensors, antiferromagnetic (AFM) systems with zero net magnetic moments are receiving much more attention due to their potential spintronic applications during the past few decades \cite{Baltz18RMP,Jungwirth16NN,Han23NM}. In spite of these discoveries, one of the main challenges for AFM materials is how to control and detect their order parameters (such as N\'eel vectors in the most abundant collinear AFMs) \cite{Nemec18NP,Wang23PRL,Shao20PRL,Xue23PRB,Han24SA,Tao24PRL}. Very recently, another type of spin antiparallel aligned systems, altermagnetic (AM) materials \cite{Smejkal22PRX,Smejkal22PRX-2,GH21PRL,Yuan20PRB,Yuan23NC,Betancourt23PRL,Zhu24Nature,Krempasky24Nature,Bai24AFM,Fedchenko24SA,Fang24PRL,Bhowal24PRX}, become an attractive topic due to their non-relativistic spin splitting feature in $\bm k$-space, protected by specific crystalline symmetry. Conventional AFMs, on the contrary, do not show finite spin splittng, at least when spin-orbit coupling (SOC) is turned off \cite{Yuan21PRM}. As both exhibit zero net magnetization, AFMs and AMs behave quite similarly in many aspects, and is potentially converted between each other under spin and orbital order competing \cite{Leeb24PRL}. It has been demonstrated that the anomalous Hall effect emerges in many (not all) AMs \cite{Feng22NE,Reichlova24NC}, while collinear AFMs do not possess a finite conductance, as in conventional nonmagnetic (NM) materials. Despite of these, more strategies are called on their distinct spintronic (in addition to transport) behaviors, which would show contrasting feature in AFMs and NMs as well.

\begin{figure}[b]
    \centering
    \includegraphics[width=0.48\textwidth]{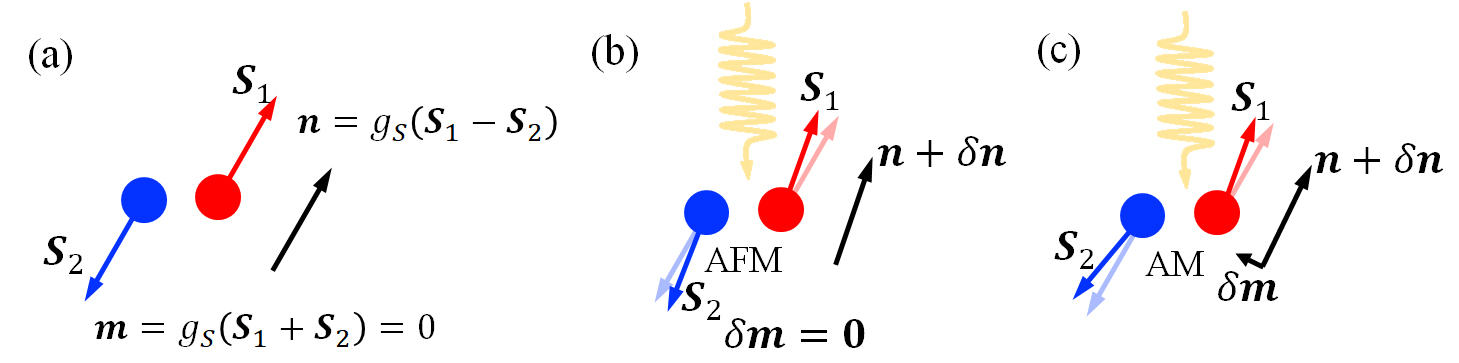}
    \caption{(a) Schematic plot of collinear antiparallel magnetic system with sizable N\'eel vector $\bm n$ and vanishing magnetic moment $\bm m$. Under LPL irradiation, the $\delta\bm n$ arises for both (b) AFMs and (c) AMs, while additional spin canting $\delta\bm m$ can be observed in the latter case. Arrows with light red and light blue colors represent their intact directions.}
    \label{fig:schematic}
\end{figure}

In this work, we predict an experimentally feasible strategy to control the N\'eel vector in collinear AFM and AM materials with compensated total magnetic moments. By scrutinizing their magnetic symmetry and the symmetry-constrained susceptibility functions, we propose that a linearly polarized light (LPL) would trigger contrasting nonequilibrium torque behaviors on the two order parameters, i.e., N\'eel vector ($\bm n=\bm m_1-\bm m_2$, with $\bm m_1$ and $\bm m_2$ referring to two antiparallel local magnetic moments) and total magnetization ($\bm m=\bm m_1+\bm m_2$) [Fig. \ref{fig:schematic}(a)]. This process is rooted by the electronic quantum metric tensor $g^{ij}$, reflecting the geometric nature of Bloch wavefunctions. Note that light-dressed magnetization responses have been widely studied over the past decades \cite{Kirilyuk10RMP,Beaurepaire96PRL,Pervishko15PRB,Kibis22PRA,Claassen17NC,Juraschek20PRR,Ren24PRL,Fei21PRL,Watzel22PRL}, serving as an ultrafast control and detection tool for spintronic devices. We apply group theory to develop the selection rules for LPL-induced torques, denoted as $\delta\bm n$ and $\delta\bm m$. Our theory indicates that for AFM materials with Kramer's degeneracy, only N\'eel vector could reorient (nonzero $\delta\bm n$ and zero $\delta\bm m$), while for AMs, both finite $\delta\bm n$ and $\delta\bm m$ arise, giving a light-induced spin canting pattern. Then we adopt simplified $\bm k\cdot\bm p$ models to explicitly show their dependence on the the equilibrium N\'eel vector $\bm n$. Finally, we perform density functional theory (DFT) calculations to quantify such processes in two realistic materials, AFM $\mathrm{FeSe}$ monolayer and AM $\mathrm{V}_2\mathrm{Se}_2\mathrm{O}$ monolayer. The magnetic moment variation could reach $10^{-2}\,\mu_B$ under intermediate light intensity, which is significant to be measured in experiments. These results are fully consistent with the symmetry arguments.

\textit{Symmetry arguments.}$-$ The main feature of AMs is non-relativistic (without SOC) spin splitting in general $\bm k$ points, while conventional AFMs exhibit spin degeneracy. We will focus on colliner spin configurations, as it is still challenging to extend the AMs into noncollinear configurations \cite{Cheong24npjQM}. According to previous works \cite{Yuan21PRM}, AMs and AFMs conceive different symmetry constraints. When the AFMs contain a combined inversion ($\mathcal{P}$) and time-reversal ($\mathcal{T}$) symmetry $\mathcal{PT}$, spin degeneracy occurs even with SOC. They belong to type-III or type-IV magnetic space groups (MSGs). Typical examples include bulk $\mathrm{CuMnAs}$ \cite{Olejnik17NC,Wang21PRL,Xu20PRB}, $\mathrm{Mn}_2\mathrm{Au}$ \cite{Bodnar18NC,Reimers23NC,AlHamdo23PRL} and $\mathrm{FeSe}$ monolayer \cite{Cao15PRB,Zhou18PRL,Wang16NM,Wang18PRL}. AFMs could also be protected by a combined $\mathcal{T}$ with a fractional translation $\bm t$, $\mathcal{T}\bm t$ (may contain $\mathcal{PT}$), such as bulk $\mathrm{MnBi}_2\mathrm{Te}_4$ \cite{Li19SA,Li23NSR} and $\mathrm{BiCoO}_3$ \cite{Yamauchi19PRB}. This could exhibit in centrosymmetric or non-centrosymmetric crystals (without considering magnetic pattern), belonging to type-IV MSG.

On the contrary, the AMs are defined in spin group theory without SOC effect \cite{Smejkal22PRX,Smejkal22PRX-2,Liu22PRX}. They possess the spin symmetry $\{\mathcal{C}_2||\mathcal{R}\}$ with $\mathcal{R}$ being either an even-ordered crystallographic rotation operator or mirror reflection (but not inversion) on the lattice, and $\mathcal{C}_2$ is perpendicular to the spin axis \cite{Smejkal22PRX,Pan24PRL}. Hence, one can easily deduce that the spin collinear AMs (with SOC, even marginal) all belong to either type-I or type-III MSGs, rather than type-IV MSGs. Typical materials include $\mathrm{CrSb}$ \cite{Reimers24NC,Zeng24AS,Ding24arxiv}, $\mathrm{MnTe}$ \cite{Osumi24PRB,Lee24PRL}, $\mathrm{MnTe}_2$ \cite{Zhu24Nature}, and $\mathrm{V}_2\mathrm{Se}_2\mathrm{O}$ \cite{Lin18PRB,Ma21NC,Jiang24arxiv,Zhang24arxiv}. There have been some debates on the spin character of $\mathrm{RuO}_2$ \cite{Shao20PRL,Feng22NE,Bai22PRL,Zhu19PRL,lin2024observationgiantspinsplitting,Liu24PRL,Kebler24npjSpin}, which awaits further investigations to explore their magnetic symmetry. Our approach provides a potential approach to characterize its ground state.

In the following, we use magnetic point group (MPG) to perform our analysis, which is sufficient to clarify and determine the light-induced total spin and N\'eel vector torque (see Supplemental Material \cite{supp} for detail). In MPG theory, the translation does not arise, and hence the type-IV MSGs (with $\mathcal{T}\bm t$) correspond to type-II gray MPGs (denoted as $\mathcal{M}_{\mathrm{II}}$). The colorless and black-white MSGs directly correspond to type-I ($\mathcal{M}_{\mathrm{I}}$) and type-III ($\mathcal{M}_{\mathrm{III}}$) MPGs, respectively. For clarity reason, we denote the $\mathcal{PT}$-invariant AFMs $\mathcal{M}_{\mathrm{III}}^{(1)}$ and those for AMs as $\mathcal{M}_{\mathrm{III}}^{(2)}$. In general, the type-III MPG is $\mathcal{M}_{\mathrm{III}}=\mathcal{D}+(\mathcal{G}-\mathcal{D})\mathcal{T}$, where $\mathcal{D}$ is a subgroup of $\mathcal{G}$ with index 2, and the coset $\mathcal{G}-\mathcal{D}=\mathcal{AD}$ with $\mathcal{A}$ being elements in $\mathcal{G}-\mathcal{D}$. Hence, $\mathcal{PT}$-AFM has $\mathcal{P}\in\mathcal{A}\mathcal{D}$. In centrosymmetric AMs (defined in nontrivial spin Laue groups), one has $\mathcal{P}\in\mathcal{D}$. Actually, in our discussion, $\mathcal{P}$ is not necessary even though most hitherto discussed AMs are centrosymmetric.

For the different MPGs as categorized previously, we analyze various symmetry-adapted susceptibility functions, in a general form $\chi_E^S=S^{0,1}E^{0,1,2}$ (see Ref. \cite{supp} for details). We suggest that the LPL-induced spin torque could trigger contrasting nonequilibrium behaviors onto the local spin sublattices, which is equivalent to the N\'eel vector torques and magnetization generation. The total magnetic torque in each unit cell is $\delta m_{c}=\tau\eta_{ab}^{S_{c}}(\omega)E_aE_b$, where $E_a$ is the alternating electric field strength and $\tau$ is the carrier lifetime. According to the Kubo perturbation theory \cite{Mukamel95Book,Sipe20PRB,Freimuch21PRB,Xu21PRB,Sturman20PU}, the susceptibility function scales with the quantum metric tensor, in the form of
\begin{equation}\label{eq:eta}
    \begin{split}
        \eta_{ab}^{S_c}(\omega)=&-\frac{g_SV_{\mathrm{u.c.}}\pi e^2}{2\hbar^2}\int [d\bm k]\delta(\omega_{mn}-\omega) \\
        &\times\sum_{m,n}f_{nm}\Delta S_{nm}^{c}\{r_{mn}^a,r_{nm}^b\}.
    \end{split}
\end{equation}
Here, we ignore the Fermi surface contribution for intrinsic semiconductors. In Eq. (\ref{eq:eta}), $g_S$ is the spin Land\'e $g$-factor, $V_{\mathrm{u.c.}}$ is the unit cell volume. One sees that the spin torque scales with quantum metric tensor $g_{mn}^{ab}=\{r_{mn}^a,r_{nm}^b\}$ ($r_{mn}^a=i\langle m|\partial_{k_a}|n\rangle$ is the interband Berry connection). $f_{mn}=f_m-f_n$ and $\Delta S_{nm}^{c}=S_{nn}^{c}-S_{mm}^{c}$ are the differences of the Fermi-Dirac distribution and spin angular momentum, respectively. All quantities are momentum-dependent, and the integral is performed over the first Brillouin zone. Note that Eq. (\ref{eq:eta}) evaluates the diagonal components of the second order responses under LPL irradiation, while the off-diagonal terms (such as spin-shift processes) are omitted as they are usually much smaller \cite{Xue24ACSNano,Xue23PRB,Fregoso22PRB,sarkar24arxiv}.

One can perform a simple symmetry analysis for Eq. (\ref{eq:eta}). As the quantum metric is invariant under $\mathcal{P}$ and $\mathcal{T}$, and $\Delta S_{nm}^{c}$ is $\mathcal{P}$ even and $\mathcal{T}$ odd, then $\eta_{ab}^{S_c}$ is $\mathcal{P}$ even, $\mathcal{T}$ odd, and $\mathcal{PT}$ odd. Hence, for AMs in $\mathcal{M}_{\mathrm{I}}$ and $\mathcal{M}_{\mathrm{III}}^{(2)}$, the LPL-induced spin torque could trigger a nonequilibrium net magnetization (finite $\delta\bm m$). In Table \ref{tab:P-symmetry}, we list all the MPGs that are $\mathcal{P}$ invariant and symmetrically-restricted zero net magnetization, with their symmetrically allowed $\eta$ components. They do not exhibit finite anomalous Hall effect, as usually conducted for AM transport detections. There are six colorless groups and five black-white groups. If we do not restrict the intrinsic zero net magnetic moments, there are additional 58 MPGs that could host finite $\eta$, which are listed in Table S2 \cite{supp}. Here, the MPGs with forbidden $\eta$ do not host collinear spin configuration, hence are out of our discussion scope. Note that in these situations, they also include reduced MPGs that allow non-relativistic spin splitting even at $\Gamma$, which does not contain $\mathcal{R}$ as defined for AMs \cite{Yuan24PRL}. In contrast, the $\eta_{ab}^{S_c}$ are to be completely silent for 21 $\mathcal{M}_{\mathrm{III}}^{(1)}$ and 32 $\mathcal{M}_{\mathrm{II}}$ groups (zero $\delta\bm m$), under symmetry constraints. Thus, we have investigated all 122 MPGs for their light-induced spin torque behaviors, and it serves as a necessary and sufficient condition to detect AMs from AFMs.

\begin{table}[t]
    \caption{Centrosymmetric MPGs with constrained vanishing intrinsic equilibrium magnetization and their allowed LPL-induced total spin torque components.}
        \centering
    \begin{tabular*}{\hsize}{@{}@{\extracolsep{\fill}}cc@{}}
    \hline\hline
     MPGs       & Allowed components       \\
    \hline
    $mmm.1$     &  $\eta_{xy}^{S_z}$, $\eta_{zx}^{S_y}$, $\eta_{yz}^{S_x}$  \\
    $4'/m$     &  $\eta_{yz}^{S_x}=\eta_{zx}^{S_y}$, $\eta_{xx}^{S_z}=-\eta_{yy}^{S_z}$, $\eta_{zx}^{S_x}=-\eta_{yz}^{S_y}$, $\eta_{xy}^{S_z}$  \\
    $4/mmm.1$    &    $\eta_{yz}^{S_x}=-\eta_{zx}^{S_y}$ \\
    $4'/mm'm$   &  $\eta_{zx}^{S_x}=-\eta_{yz}^{S_y}$,  $\eta_{xx}^{S_z}=-\eta_{yy}^{S_z}$ \\
    $\bar{3}m.1$   &  $\eta_{xx}^{S_x}=-\eta_{yy}^{S_x}=-\eta_{xy}^{S_y}$, $\eta_{yz}^{S_x}=-\eta_{zx}^{S_y}$  \\
    $6'/m'$     &  $\eta_{xx}^{S_x}=-\eta_{yy}^{S_x}=-\eta_{xy}^{S_y}$, $\eta_{xy}^{S_x}=\eta_{xx}^{S_y}=-\eta_{yy}^{S_y}$  \\
    $6/mmm.1$     &  $\eta_{yz}^{S_x}=-\eta_{zx}^{S_y}$  \\
    $6'/m'mm'$     &  $\eta_{xx}^{S_x}=-\eta_{yy}^{S_x}=-\eta_{xy}^{S_y}$  \\
    $m\bar{3}.1$     &  $\eta_{yz}^{S_x}=\eta_{zx}^{S_y}=\eta_{xy}^{S_z}$  \\
    $m\bar{3}m.1$ & None \\
    $m\bar{3}m'$     &  $\eta_{yz}^{S_x}=\eta_{zx}^{S_y}=\eta_{xy}^{S_z}$  \\
    \hline\hline
    
    \end{tabular*}
    \label{tab:P-symmetry}
\end{table}

We argue that the collinear AFM with $\mathcal{M}_{\mathrm{III}}^{(1)}$ and $\mathcal{M}_{\mathrm{II}}$ could show magnetic sector (or sublattice) dependent spin torque under light irradiation. Here, each sector contains parallel spin polarization, as depicted in Fig. \ref{fig:schematic}(b). This can be evaluated by introducing a N\'eel operator $\hat{\bm L}=\hat{\bm S}_{\alpha}-\hat{\bm S}_{\beta}$ ($\alpha$ and $\beta$ are two antiparallel spin polarizations), and replace the $\Delta S_{nm}^c$ operator in Eq. (\ref{eq:eta}) by $\Delta L_{nm}^c$. Then, we denote the susceptibility as $\zeta_{ab}^{L_c}(\omega)=\eta_{ab}^{S_{\alpha,c}}(\omega)-\eta_{ab}^{S_{\beta,c}}(\omega)$, which indicates a finite N\'eel vector torque $\delta\bm n$.

As there is no direct symmetry arguments for $\hat{\bm L}$, we deduce a general magnetic group analysis for the symmetry constraints on $\zeta_{ab}^{L_c}(\omega)$. It is evident that this is equivalent to sector-dependent local and hidden spin torques, and it would follow a MPG (denoted as $\mathcal{N}$) that is a subgroup of the original MPG. Hence, we examine all the operators that are locally compatible within the sector. We denote operators $\mathcal{O}_1$ (and $\mathcal{O}_2$) that transform the sites between (and within) the antiparallel sectors,
\begin{equation}
    \mathcal{O}_1\bm r_i^{\alpha}=\bm r_j^{\beta},\qquad\mathcal{O}_1\bm m_i^{\alpha}=\bm m_j^{\beta}=-\bm m_i^{\alpha},
\end{equation}
and
\begin{equation}
    \mathcal{O}_2\bm r_i^{\alpha}=\bm r_j^{\alpha},\qquad\mathcal{O}_2\bm m_i^{\alpha}=\bm m_j^{\alpha}.
\end{equation}
Here, $i$ and $j$ are site indices. In MSG representations without SOC \cite{Yuan24PRL}, they correspond to primed and unprimed operators, while with SOC it depends on the spin (N\'eel) direction. For the $\mathcal{M}_{\mathrm{III}}^{(1)}$ MPGs, it is clear that the inversion symmetry operator $\mathcal{P}\in\{\mathcal{O}_1\}$. If the operator $\mathcal{O}_1\in\mathcal{D}$, for spin angular momentum $S_c$, it will flip their sign, namely, the spin is variant under such an operator. This can be reflected by the character of the irreducible representation of rotation basis set (that transforms the same as spin angular momentum) in point group $\mathcal{G}$, giving $\chi_{\Gamma_{S_c}}({\mathcal{O}_1})=-1$. Note that for collinear AFM systems, no degenerate irreducible representations are allowed. Similarly, it can be shown that if $\mathcal{O}_2\in\mathcal{D}$, it maintains the spin direction with $\chi_{\Gamma_{S_c}}({\mathcal{O}_1})=1$. On the contrary, if $\mathcal{O}_1\in\mathcal{G}-\mathcal{D}$, we have $\chi_{\Gamma_{S_c}}({\mathcal{O}_1})=1$; and $\chi_{\Gamma_{S_c}}({\mathcal{O}_2})=-1$ if $\mathcal{O}_2\in\mathcal{G}-\mathcal{D}$.

In order to determine whether the intra-sector operator $\mathcal{O}_2$ remains to be invariant, we adopt isomorphic group method with respect to the Birss's notation \cite{Birss64Book}. This requires to use an irreducible representation $\Gamma_m$ of the MPG. The $\Gamma_m$ keeps the operators in $\mathcal{D}$ to be invariant [positive $\chi_{\Gamma_m}({\mathcal{O}_{\mathcal{D}}})$] while reverses the operators in $\mathcal{G}-\mathcal{D}$ [negative $\chi_{\Gamma_m}({\mathcal{O}_{\mathcal{G}-\mathcal{D}}})$] \cite{Wang20npjCM}. Therefore, the $\mathcal{O}_2$ would remain to be invariant in $\mathcal{N}$ if $\chi_{\Gamma_m}({\mathcal{O}_2})$ and $\chi_{\Gamma_{S_c}}({\mathcal{O}_2})$ have the same sign. The set of such operators form $\mathcal{N}$, which is clearly the subgroup of $\mathcal{M}_{\mathrm{III}}^{(1)}$. For example, if we take the MPG of $2'/m$ that is $\mathcal{PT}$ invariant, the $\Gamma_m$ of point group $2/m$ is $B_u$. The spin operators transform as $\Gamma_{S_x}=\Gamma_{S_y}=B_g$ and $\Gamma_{S_z}=A_g$. By comparing their signs (Table S3 \cite{supp}), it is clearly that the spin-$x$ and $y$ follow MPG $2'$, and the spin-$z$ follows $m.1$. Performing the aforementioned analysis, the $2'$ MPG (taking the axis along $z$) could yield finite $\zeta_{xx}^{L_x}$, $\zeta_{yy}^{L_x}$, $\zeta_{zz}^{L_x}$, $\zeta_{xy}^{L_x}$, $\zeta_{xx}^{L_y}$, $\zeta_{yy}^{L_y}$, $\zeta_{zz}^{L_y}$, $\zeta_{xy}^{L_y}$, $\zeta_{yz}^{L_z}$, and $\zeta_{zx}^{L_z}$. The $m.1$ MPG (mirror vertical to $z$), on the other hand, gives finite $\zeta_{xx}^{L_z}$, $\zeta_{yy}^{L_z}$, $\zeta_{zz}^{L_z}$, $\zeta_{xy}^{L_z}$, $\zeta_{zx}^{L_x}$, $\zeta_{zx}^{L_y}$, $\zeta_{yz}^{L_x}$, and $\zeta_{yz}^{L_y}$. Hence, if the light polarization is off the principal axis (such as $xy$), one achieves N\'eel vector torques along $x$ and $y$ (away from the equilibrium N\'eel direction). In both cases, the components are independent.

We scrutinize all the $\mathcal{PT}$-invariant $\mathcal{M}_{\mathrm{III}}^{(1)}$, and list their corresponding $\mathcal{N}$ for each N\'eel vector component in Table \ref{tab:PT-subgroup}. Clearly, the missing operators from $\mathcal{M}_{\mathrm{III}}^{(1)}$ to $\mathcal{N}$ form set of $\mathcal{O}_1$, so that the $\mathcal{N}$'s are not $\mathcal{PT}$ invariant. For all 32 gray $\mathcal{M}_{\mathrm{II}}$ MPGs, we list their subgroups $\mathcal{N}$ in Table S4 \cite{supp}. They do not contain either $\mathcal{PT}$ or $\mathcal{T}$. Hence, they exhibit nonzero N\'eel vector torques under LPL irradiation. Note that the AMs with finite $\delta\bm m$ would also give finite $\delta\bm n$, as the latter one follows a subgroup symmetry argument of the former [Fig. \ref{fig:schematic}(c)].

\begin{table}[t]
    \caption{Subgroup $\mathcal{N}$ for each N\'eel vector components ($L_x,L_y,L_z$) in parity-time $\mathcal{PT}$ invariant MPGs, which can be used to determine their $\delta\bm n$, with the $\delta\bm m$ always vanishing. The \xmark \,indicates that a unidirectional N\'eel vector is not allowed.}
        \centering
    \begin{tabular*}{\hsize}{@{}@{\extracolsep{\fill}}cccc@{}}
    \hline\hline
     $\mathcal{M}_{\mathrm{III}}^{(1)}$ MPGs    & $\mathcal{N}(L_x)$  &   $\mathcal{N}(L_y)$  &      $\mathcal{N}(L_z)$     \\
    \hline
    $\bar{1}'$  &  $1.1$  &  $1.1$  & $1.1$  \\
    $2/m'$  &  $m'$  &  $m'$  & $2.1$  \\
    $2'/m$  &  $2'$  &  $2'$  & $m.1$  \\
    $m'mm$  &  $m'm2'$  &  $m'm2'$  & $2'2'2$  \\
    $m'm'm'$  &  $m'm'2$  &  $m'm'2$  & $m'm'2$  \\
    $4/m'$  &  \xmark  &  \xmark  & $4.1$  \\
    $4'/m'$  &  \xmark  &  \xmark  & $\bar{4}.1$  \\
    $4/m'mm$  &  \xmark &  \xmark  & $42'2'$  \\
    $4'/m'm'm$  &  \xmark  &  \xmark  & $\bar{4}2'm'$  \\
    $4/m'm'm'$  &  \xmark  &  \xmark  & $4m'm'$  \\
    $\bar{3}'$  &  \xmark &  \xmark  & $3.1$  \\
    $\bar{3}'m$  &  \xmark  &  \xmark  & $32'$  \\
    $\bar{3}'m'$  &  \xmark  &  \xmark  & $3m'$  \\
    $6'/m$  &  \xmark  &  \xmark  & $\bar{6}.1$  \\
    $6/m'$  &  \xmark  &  \xmark  & $6.1$  \\
    $6/m'mm$  &  \xmark  &  \xmark  & $62'2'$  \\
    $6'/mmm'$  & \xmark  &  \xmark  & $\bar{6}m2.1$  \\
    $6/m'm'm'$  &  \xmark  &  \xmark  & $6m'm'$  \\
    $m'\bar{3}'$  &  \xmark  &  \xmark  & \xmark \\
    $m'\bar{3}'m$  &  \xmark  &  \xmark  & \xmark  \\
    $m'\bar{3}'m'$  &  \xmark  &  \xmark  & \xmark \\
    \hline\hline
    
    \end{tabular*}
    \label{tab:PT-subgroup}
    \end{table}

\textit{Low energy $\bm k\cdot\bm p$ model.}$-$ Having these group theory analyses, we now take two typical low energy Hamiltonian models to explicitly conduct the LPL-induced spin torque. For the $\mathcal{PT}$ invariant AFM system, we take a representative $4\times 4$ model on a square lattice \cite{Smejkal17PRL}
\begin{equation}\label{eq:afm}
    \begin{split}
        H_{\mathrm{AFM}}=&-2t\cos\frac{k_x}{2}\cos\frac{k_y}{2}\tau_x+J_{\mathrm{ex}}\tau_z\sigma_{\bm n} \\
        &+\lambda_{\mathrm{SOC}}(\sin k_x\tau_z\sigma_y-\sin k_y\tau_z\sigma_x).
    \end{split}
\end{equation}
Here, $t$ refers to inter-site interaction, $J_{\mathrm{ex}}$ is the exchange energy splitting along N\'eel vector direction $\bm n$, and $\lambda_{\mathrm{SOC}}$ is SOC interaction in the Rashba form. $\bm \tau$ and $\bm \sigma$ denote Pauli matrices for the orbital (sublattice) and spin degrees of freedom, respectively. $\sigma_{\bm n}=\bm n\cdot\bm\sigma$ is along the N\'eel vector direction. The typical band dispersion can be found in Fig. S1 \cite{supp}. Note that in this model, the equilibrium N\'eel vector operator can be written as $\tau_z\sigma_{\bm n}$.

\begin{figure}[t]
    \centering
    \includegraphics[width=0.48\textwidth]{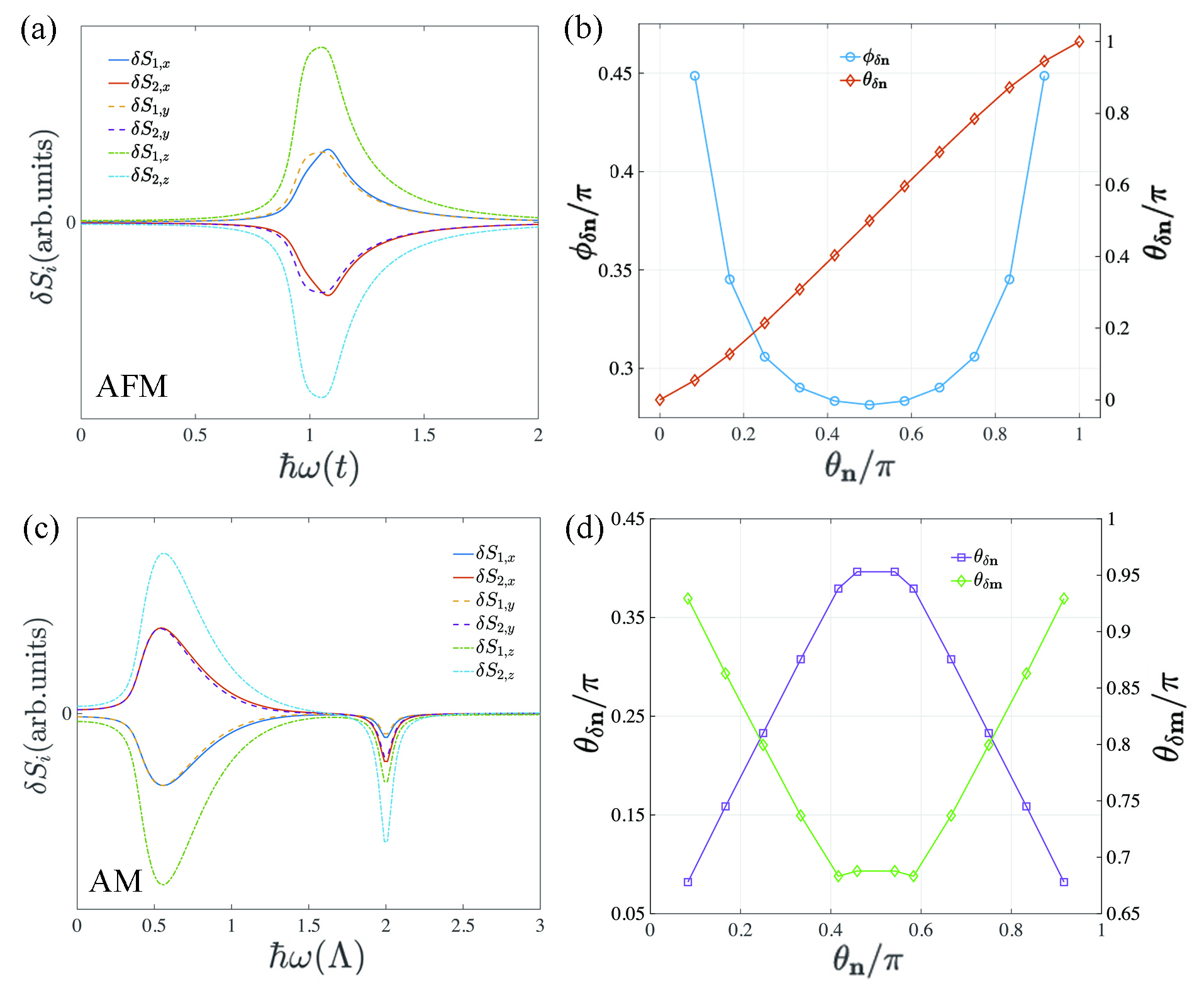}
    \caption{Low energy model calculation results. (a) Sublattice dependent variation of spin torque components in the $\mathcal{PT}$ square lattice. The spherical angles of equilibrium N\'eel vector is $\theta_{\bm n}=\pi/6$ and $\phi_{\bm n}=\pi/4$.  The model parameters in Eq. (\ref{eq:afm}) are $J_{\mathrm{ex}}=0.5t$ and $\lambda_{\mathrm{SOC}}=0.2t$. (b) The N\'eel vector torque directions $\theta_{\delta\bm n}$ and $\phi_{\delta\bm n}$ as functions of the $\theta_{\bm n}$, with the incident photon energy fixed at $\hbar\omega=1.0t$ and the azimuthal angle of equilibrium N\'eel vector at $\phi_{\bm n}=\pi/4$. (c) The spin torque components with in a AM model, with the equilibrium N\'eel vector along $\theta_{\bm n}=\pi/6$ and $\phi_{\bm n}=\pi/4$. The model parameters in Eq. (\ref{eq:am}) are $J_{\mathrm{ex}}=0.4\Lambda$ and $\lambda_{\mathrm{SOC}}=0.2\Lambda$. (d) The polar angles of N\'eel vector torque $\delta\bm n$ and induced magnetic moment $\delta\bm m$ as functions of the equilibrium N\'eel vector polar angle $\theta_{\bm n}$, with a fixed $\phi_{\bm n}=\pi/4$. The incident photon energy is chosen to be $\hbar\omega=2.0\Lambda$. In all calculations, $x$-polarized incident light is assumed.}
    \label{fig:kpmodel}
\end{figure}

Figure \ref{fig:kpmodel}(a) shows the LPL-induced sublattice-dependent spin torque for each spin component. One sees that these torques are exactly antiparallel on the two sublattices, giving only finite $\delta\bm n$ with vanishing $\delta\bm m$. When the equilibrium N\'eel vector rotates from $+z$ to $-z$ directions (with fixed azimuthal angle $\phi_{\bm n}=\pi/4$), the N\'eel torque directions ($\phi_{\delta\bm n}$ and $\theta_{\delta\bm n}$) exhibit nonlinear relationships with respect to $\theta_{\bm n}$. When the equilibrium N\'eel vector is away from the $z$ axis or $xy$ plane, the N\'eel torque $\delta\bm n$ will be away from the $\bm n$. Here the azimuthal angle $\phi_{\delta\bm n}$ is significantly varied from its equilibrium direction $\phi_{\bm n}=\pi/4$. The polar angle of N\'eel torque also deviates from the equilibrium state. These results are fully consistent with previous group theory results. We find that such a N\'eel torque effect can be enhanced by increasing the SOC strength (Fig. S2 \cite{supp}). In order to elucidate the quantum geometry feature, we plot the optical quantum metric $\sum_{c,v,\bm k}g_{cv}^{xx}\delta(\omega_{cv}-\omega)$ under $x$-LPL in Fig. S3 \cite{supp}. Note that the quantum metric tensor has been receiving tremendous attention recently \cite{Gao23Science,Wang23Nature,Han24NatPhys,Kang25NatPhys}. Similar results can be obtained for honeycomb $\mathcal{PT}$ invariant lattices (see SM for more details \cite{supp}).

As for the AMs, we take a typical simplified low energy Hamiltonian also in a square lattice \cite{Zhang24PRL}
\begin{equation}\label{eq:am}
    \begin{split}
        H_{\mathrm{AM}}=&\Lambda\tau_z+J_{\mathrm{ex}}\tau_z\sigma_{\bm n}\cos k_y\cos k_x \\
        +&\lambda_{\mathrm{SOC}}\tau_z(\sigma_y\sin k_x-\sigma_x\sin k_y).
    \end{split}
\end{equation}
Here, $\Lambda$ measures the on-site energy difference between orbitals. The band dispersion, optical quantum metric, and its SOC-dependence are plotted in Figs. S4-S6 \cite{supp}. Figure \ref{fig:kpmodel}(c) illustrates the spin torque components for the two magnetic sublattices. It is evident that the total spin torques (summed over the two sublattices) do not vanish, consistent with our previous symmetry arguments. Hence, LPL could apply a spin canting into AMs, giving finite $\delta\bm m$ as well as $\delta\bm n$. Figure \ref{fig:kpmodel}(d) shows how the polar angle of the equilibrium N\'eel vector controls the polar angles of $\delta\bm m$ and $\delta\bm n$ for a specific incident energy ($2.0\Lambda$) that is slightly above the overall bandgap. In this model, the $\theta_{\delta\bm n}$ always along the upper side of the hemisphere ($<0.5\pi$), while the spin canting related $\theta_{\delta\bm m}$ is pointing along the downward ($>0.5\pi$).

\begin{figure}[t]
    \centering
    \includegraphics[width=0.47\textwidth]{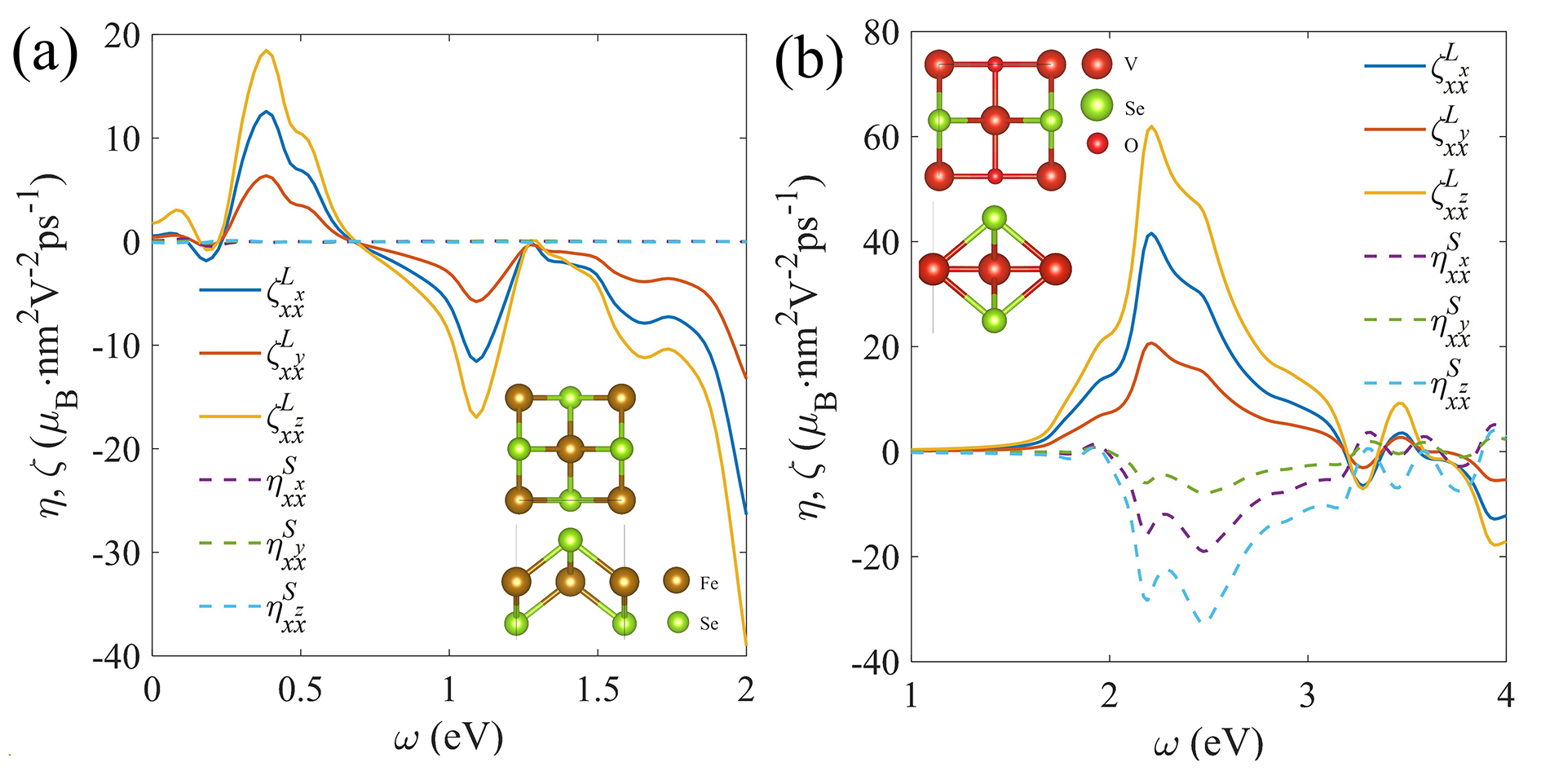}
    \caption{DFT calculation results. Top and side view of (a) AFM $\mathrm{FeSe}$ monolayer and its spin and N\'eel torque components. (b) is the corresponding results for the AM $\mathrm{V}_2\mathrm{Se}_2\mathrm{O}$ monolayer. The spherical angles of equilibrium N\'eel vector are set as general values of $\theta_{\bm n}=0.204\pi$ and $\phi_{\bm n}=0.148\pi$. In these results, $x$-polarized light is applied.}
    \label{fig:DFT}
\end{figure}

\textit{Realistic materials.}$-$ We apply \textit{ab initio} calculations to illustrate the contrasting LPL-induced spin torque responses using two realistic materials, namely, a $\mathrm{FeSe}$ monolayer [Fig. \ref{fig:DFT}(a)] and a $\mathrm{V}_2\mathrm{Se}_2\mathrm{O}$ monolayer [Fig. \ref{fig:DFT}(b)]. The former one is a typical AFM material with $\mathcal{PT}$ symmetry (exhibiting high $T_c$ Fe-based superconducting feature), and the latter one has a spatial protected valley polarization, belonging to $\mathcal{P}$ symmetric $d$-wave AM systems. Their band structures are given in Fig. S9 \cite{supp}. Consistent with the low-energy model, for AFM $\mathrm{FeSe}$, the total $\eta$ is always zero and one has finite $\zeta$. The detailed symmetry arguments can be seen in SM \cite{supp}. For instance, if the incident photon energy is $\hbar\omega=0.28\,\mathrm{eV}$, the $\zeta$ could reach $~10\,\frac{\mu_B\mathrm{nm}^2}{\mathrm{V}^2\mathrm{ps}}$. Hence, when the light alternating electric field strength is $E=0.1\,\mathrm{V}/\mathrm{nm}$, the N\'eel vector moment variation could reach $1\times 10^{-2}\,\mu_B$ if one assumes an conservative and experimentally achievable $\tau=0.1\,\mathrm{ps}$. This $\delta\bm n$ is along $\phi_{\delta\bm n}=0.157\pi$ and $\theta_{\delta\bm n}=0.199\pi$, which is sufficiently large to be measured in experiments (giving about $2^{\circ}$ difference from the equilibrium N\'eel vector). As a second order optical process, enhancing light intensity linearly increases $\delta\bm n$.

On the contrary, for the AM $\mathrm{V}_2\mathrm{Se}_2\mathrm{O}$ monolayer, the calculated results suggest both finite $\delta\bm m$ and $\delta\bm n$. At the incident photon energy of $\hbar\omega=2.2\,\mathrm{eV}$, the total induced magnetic moment could reach $0.03\,\mu_B$ under $E=0.1\,\mathrm{V}/\mathrm{nm}$, along $\phi_{\delta\bm m}=1.117\pi$ and $\theta_{\delta\bm m}=0.833\pi$. This indicates a detectable spin canting in the unit cell. For both cases, we summarize their N\'eel vector dependent results in Fig. S10 \cite{supp}. The $\bm k$-space dependent spin torque contributions in both materials are also shown in Fig. S11 \cite{supp}.

\textit{Discussion and Conclusion.}$-$ In this work, we evaluate the spin polarization contributed magnetic moments, and ignore the orbital moment contributions. In spin polarized systems, the SOC effect constrains that the orbital moments follow the same symmetry arguments with spin. Hence, our MPG results hold for orbital degree of freedom. In addition, even recent studies have shown large orbital moments during nonequilibrium process \cite{Go20PRResearch,Cysne21PRL,Jo24npjSpin}, they usually dominate in non-spin polarized or paramagnetic materials. For spin polarized systems, the spin order still dominates.
For noncollinear AFMs such as many Kagom\'e lattice based materials, one can obtain local spin torques on each spin sublattice. Our sublattice dependent hidden spin torque can be viewed as dividing the whole AFM systems into parallel spin sectors. Recent works have proposed that AFMs could show non-relativistic spin splitting in local sectors containing antiparallel spin configuration \cite{Yuan23NC}. If such sector selections are adopted (such as $C$ or $G$ type layered AFMs), one can still anticipate net light-induced magnetization occurs within each sector, as both $\mathcal{PT}$ and $\mathcal{T}\bm t$ symmetries are locally broken. These are beyond the scope of our current work, and will be discussed separately. Unlike anomalous Hall effect that is silent in both collinear AFMs and NMs, the N\'eel vector torque is clearly vanishing for NMs. Furthermore, while the anomalous Hall effect only exhibits in a subset of AMs, this light-induced (global and local) spin torque, protected by group theory, could serve as a powerful tool to distinguish and disentangle AFMs, AMs, and NMs.

In conclusion, we develop a group theory method to elucidate how LPL induces local spin torque on the magnetic sites in collinear AFM and AM materials. The former type, protected by $\mathcal{PT}$ or $\mathcal{T}\bm t$, only exhibits torques on the N\'eel vector, which is kept to be antiparallel. Note that the N\'eel vector rotation can be experimentally measured by its induced electric current signal \cite{Godinho18NC,Huang24NC}. The light-induced spin effect in AMs, on the contrary, could break its collinear spin polarization and show nonzero magnetic moments. Hence, they can be disentangled upon magnetic measurements. The symmetric arguments are further verified by the low energy $\bm k\cdot\bm p$ model and DFT calculations in two representative systems. Our work provides a subtle approach to disclose the symmetry difference and nonlinear optical responses among the AM and conventional AFM systems, with and without non-relativistic spin splitting, respectively.

\begin{acknowledgments}
This work is supported by the National Natural Science Foundation of China (NSFC) under Grant Nos. 12374065 and 12274342.
\end{acknowledgments}

\providecommand{\noopsort}[1]{}\providecommand{\singleletter}[1]{#1}%

\end{document}